# Ultrafast Laser Ablation, Intrinsic Threshold, and Nanopatterning of Monolayer Molybdenum Disulfide

**Joel M. Solomon[1], Sabeeh Irfan Ahmad[1], Arpit Dave[1], Li-Syuan Lu[2,3], Fatemeh HadavandMirzaee[1], Shih-Chu Lin[2], Sih-Hua Chen[2], Chih-Wei Luo[2,4,5], Wen-Hao Chang[2,3], and Tsing-Hua Her[1,\*]**

[1] Department of Physics and Optical Science, The University of North Carolina at Charlotte, Charlotte, North Carolina 28223, United States
[2] Department of Electrophysics, National Yang Ming Chiao Tung University, Hsinchu 30010, Taiwan
[3] Research Center for Applied Sciences, Academia Sinica, Taipei 11529, Taiwan
[4] Institute of Physics and Center for Emergent Functional Matter Science, National Yang Ming Chiao Tung University, Hsinchu 30010, Taiwan
[5] National Synchrotron Radiation Research Center (NSRRC), Hsinchu 30076, Taiwan
[\*] ther@uncc.edu

## ABSTRACT

Laser direct writing is an attractive method for patterning 2D materials without contamination. Literature shows that the femtosecond ablation threshold of graphene across substrates varies by an order of magnitude. Some attribute it to the thermal coupling to the substrates, but it remains by and large an open question. For the first time the effect of substrates on femtosecond ablation of 2D materials is studied using $MoS_2$ as an example. We show unambiguously that femtosecond ablation of $MoS_2$ is an adiabatic process with negligible heat transfer to the substrates. The observed threshold variation is due to the etalon effect which was not identified before for the laser ablation of 2D materials. Subsequently, an intrinsic ablation threshold is proposed as a true threshold parameter for 2D materials. Additionally, we demonstrate for the first time femtosecond laser patterning of monolayer $MoS_2$ with sub-micron resolution and mm/s speed. Moreover, engineered substrates are shown to enhance the ablation efficiency, enabling patterning with low-power femtosecond oscillators. Finally, a zero-thickness approximation is introduced to predict the field enhancement with simple analytical expressions. Our work clarifies the role of substrates on ablation and firmly establishes femtosecond laser ablation as a viable route to pattern 2D materials.

**INTRODUCTION**

Single atomic layer materials such as graphene, transition metal dichalcogenides (TMDs), and hexagonal boron nitride have been studied extensively for their novel electronic and optical properties and for their applications in optoelectronic devices.[1,2] Graphene exhibits strong wavelength-independent absorption of 2.3% and high carrier mobilities reaching 200000 cm$^2$/V s.[1,3] TMDs such as Molybdenum disulfide ($MoS_2$) and Tungsten disulfide ($WS_2$) are of great interest because of their transition from indirect to direct band gap and strong excitonic resonances at room temperature as the number of layers is reduced to a monolayer.[4,5] Both graphene and $MoS_2$ have demonstrated phenomenal mechanical robustness[6,7] and optical stability under intense femtosecond excitation.[8,9] These properties have led to the research and development of 2D material-based electronic and optoelectronic devices such as transistors,[1,10] photodetectors,[1,11] and additional heterostructure devices.[12]

For such device applications, reliable patterning techniques are essential to selectively remove 2D materials for specific sizes and geometries. Although electron beam and photolithography have been used extensively to pattern 2D materials, they suffer from high costs, complexity, vacuum operation requirements, and more importantly are prone to leave behind contaminates or polymer residues, causing damage or unwanted doping which can inadvertently degrade their electrical properties.[13] In this regard, laser ablation is a promising technique to pattern 2D materials that is in situ, resist-free, and maskless. Specifically, the ultrafast laser ablation and patterning of graphene based on oxidative burning has been demonstrated on several substrates.[14] Scanning rates as high as tens of mm/s can be achieved with a laser fluence of a couple hundred mJ/cm$^2$ from laser amplifiers.[15] In addition, sub-diffraction-limited ablated features under 100 nm can be obtained with shaped picosecond laser beams.[14] In contrast, little research has been conducted on the femtosecond ablation of TMDs. Paradisanos *et al.* has studied the multi-shot degradation of exfoliated monolayer and bulk MoS$_2$ and reported single-shot ablation thresholds based on the appearance of submicron-sized distortion.[16] Pan *et al.* investigated ablation mechanisms of bulk MoS$_2$ under intense femtosecond excitation, and determined that the ablation was mediated by sublimation at weak pumping and melting at strong pumping.[17] Despite these efforts, a rigorous investigation of the threshold fluence and ultrafast laser patterning of monolayer TMDs has not been demonstrated. We note that continuous-wave 532 nm (CW) lasers have been demonstrated to sublimate monolayer MoS$_2$ on a SiO$_2$/Si substrate with a 200 nm spatial resolution,[18] though the patterning speed is slow due to its photothermal nature. The throughput, however, can be substantially increased using a photothermoplasmonic substrate which then requires transfer of the patterned MoS$_2$ to other susbtrates.[19]

Since many applications require a supporting substrate, understanding its effect on the laser ablation of 2D materials is important. Although ultrafast laser ablation of graphene has been extensively studied, the role of the substrates is still not clear. The reported ablation thresholds from many studies made by similar pulse widths (~50-100 fs) and wavelengths (~800 nm) differ by one order of magnitude among suspended graphene and graphene supported by borosilicate glass, Al$_2$O$_3$, and 285 nm SiO$_2$/Si substrates.[15,20-24] Surprisingly, such differences have never been discussed or understood. Beyond mechanical support, substrates have been routinely claimed to act as a heat sink to explain why CW laser thinning of multi-layer graphene and MoS$_2$ self-terminates at monolayers.[18,25] Other groups also observed that the ablation threshold for both femtosecond and CW excitation are lower for suspended 2D materials than those supported on a $SiO_2$/Si substrate, which was again attributed to heat dissipation through the supporting substrates.[21,26] Optically, substrates are known to enhance the light outcoupling of 2D materials through the etalon effect. For $SiO_2$/Si substrates, the Raman scattering was shown to strongly depend on the $SiO_2$ thickness for graphene,[27] which led to the optimization of both the Raman scattering and photoluminescence of $WSe_2$ by controlling the $SiO_2$ layer thickness where the largest enhancement occurred for a $SiO_2$ thickness of about 90 nm for 532 nm excitation.[28] Similar enhancement for Raman scattering, photoluminescence, and second harmonic generation was obtained by using distributed Bragg reflectors (DBR) as a substrate for $MoS_2$.[29] Improved optical contrast of graphene and $MoS_2$ was achieved by designing multilayer heterostructure substrates where an optical contrast of 430% was obtained for monolayer $MoS_2$.[30,31] We note that the etalon effect has been previously shown to modulate the laser thinning efficiency of multilayer graphene,[25] but has never been studied for laser ablation of 2D materials.

In this work, we studied the femtosecond laser ablation of monolayer $MoS_2$ on a variety of common substrates. Notably, we demonstrated this process is both high speed (~5 mm/sec) and high resolution (~ 250 nm with a 0.55 NA objective at 800 nm). Moreover, the influence of substrates on the ablation threshold fluence $F_{th}$ was investigated, both in single-shot and line-scan modes. It was shown that femtosecond laser ablation of transferred monolayer $MoS_2$ is adiabatic where the heat dissipation through the supporting substrates is negligible, and the variation in $F_{th}$ among substrates can be largely explained by the substrates' etalon effect. Based on our finding, an all-dielectric DBR substrate was realized to reduce $F_{th}$ by 7× compared to that of sapphire to enable laser patterning using low-power femtosecond oscillators. Furthermore, we introduced an intrinsic ablation threshold fluence $F_{th}^{int}$ as a substrate-independent threshold parameter for the laser ablation of 2D materials. We also introduced the zero-thickness approximation to substantially simplify the calculation of the etalon effect for laser ablation. Combined with the knowledge of $F_{th}^{int}$, this makes incident $F_{th}$ on any substrates predictable. Our work clarifies the role of substrates and provides a foundation for rapid prototyping of 2D-material devices using femtosecond laser ablation.

**RESULTS AND DISCUSSION**

**Zero-Thickness Approximation.** Previous studies on the etalon effect of monolayer 2D materials focused on engineering the Raman scattering, photoluminescence, and second-harmonic generation by optimizing the internal field at the excitation wavelength and the outcoupling efficiency at the emission wavelength.[27-29] As a result, the theoretical enhancement can only be calculated computationally. For ablation, only the excitation enhancement is of concern and we show below that the internal field $\mathcal{E}_{2DM}$ at the excitation wavelength has a simple analytical approximation. The substrates used in this study include sapphire ($Al_2O_3$), borosilicate glass, 70 nm thick gold (Au) film on a glass substrate, 90 nm $SiO_2$/Si, and two custom designed DBR substrates: one DBR substrate (DBR800(+)) targets maximal intensity enhancement and the other (DBR800(-)) targets maximal intensity suppression. The system can be modeled as an asymmetric etalon composed of air, a 2D material, and a substrate (Figure S1a). If the effective reflection coefficient between the monolayer and the substrate $\tilde{r}_{1s} = r_o \exp(i\phi)$ is known, then the spatial distribution of the electric field inside the monolayer $\mathcal{E}_{2DM}(x)$ can be rigorously calculated using the Airy formula (see equation (S1) in the supplemental information (SI)). Since monolayer 2D materials are much thinner compared to the wavelength investigated here, we introduce the zero-thickness approximation (ZTA) to simplify the internal field $\mathcal{E}_{2DM}(x)$ to become

$$\mathcal{E}_{2DM}(x) \approx \mathcal{E}_{2DM}^{ZTA} = \mathcal{E}_{inc}\tilde{t}_{01}\left(\frac{1+\tilde{r}_{1s}}{1-\tilde{r}_{1s}\tilde{r}_{10}}\right), \quad (1)$$

where $\mathcal{E}_{inc}$ is the incident electric field and $\tilde{t}_{ij}$ and $\tilde{r}_{ij}$ are Fresnel transmission and reflection coefficients from the $i^{th}$ to $j^{th}$ medium, respectively. For single-material substrates such as $Al_2O_3$, glass, or a thick Au film, $\tilde{r}_{1s}$ is simply the Fresnel reflection coefficient (see equation (S3)), and the internal field $\mathcal{E}_{2DM}$ becomes

$$\mathcal{E}_{2DM}^{ZTA} = \mathcal{E}_{inc}\left(\frac{2}{1+\tilde{n}_s}\right), \quad (2)$$

where $\tilde{n}_s$ is the complex refractive index of the substrate. For $SiO_2$/Si substrates with a silica layer thickness of $d_2$, $\tilde{r}_{1s}$ can be calculated analytically using an asymmetric etalon composed of a TMD, $SiO_2$, and Si (see equation (S4)), and $\mathcal{E}_{2DM}$ becomes approximately

$$\mathcal{E}_{2DM}^{ZTA} = \mathcal{E}_{inc}\left(\frac{2[(\tilde{n}_2+\tilde{n}_s)-(\tilde{n}_s-\tilde{n}_2)e^{i2\beta_2 d_2}]}{(1+\tilde{n}_2)(\tilde{n}_2+\tilde{n}_s)+(\tilde{n}_2-1)(\tilde{n}_s-\tilde{n}_2)e^{i2\beta_2 d_2}}\right), \quad (3)$$

where $\beta_2 = 2\pi\tilde{n}_2/\lambda_0$, and $\tilde{n}_2$ and $\tilde{n}_s$ are the refractive indices of $SiO_2$ and Si, respectively. For DBR substrates, an analytical expression of $\mathcal{E}_{2DM}^{ZTA}$ can be found in equation (S12). Details about the DBR design and fabrication can be found in the supplemental information (Figure S2). Equations (2)-(3) and equation (S12) clearly show $\mathcal{E}_{2DM}^{ZTA}$ is independent of the 2D materials. In fact, this result can be extended to arbitrary stratified substrates with the proof being presented in the SI (see equation (S9)).

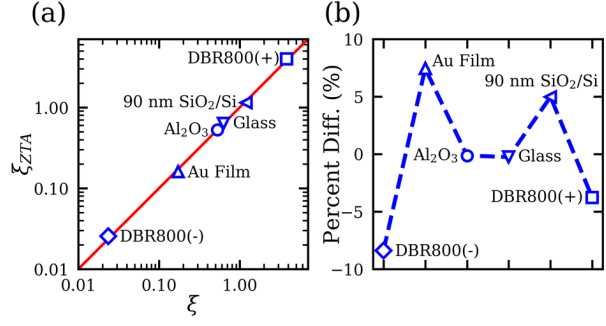

**Figure 1.** (a) Comparison of the internal intensity enhancement factor calculated from the rigorous Airy formula $\xi$ and ZTA $\xi_{ZTA}$ at 800 nm. The red line represents the ideal one-to-one ratio. (b) The percent difference between $\xi$ and $\xi_{ZTA}$ for the substrates in (a). A positive percentage means $\xi$ is larger than $\xi_{ZTA}$.

Based on equations (1)-(3), we can define an internal intensity enhancement factor $\xi = |\mathcal{E}_{2DM}|^2/|\mathcal{E}_{inc}|^2$. Figure 1a compares the $\xi_{ZTA}$ based on the ZTA with the rigorous $\xi$ calculated from equation (S1). For the latter, the intensity is averaged over the thickness of the 2D material according to equation (S2). Figure 1a shows excellent agreement between $\xi$ and $\xi_{ZTA}$ for the various substrates under consideration. The red line in Figure 1a represents the ideal one-to-one ratio. Interestingly, the 90 nm SiO$_2$/Si substrate has a $\xi$ close to unity (~1.14). Figure 1b shows that the percent differences for various substrates are all within 5% except the Au film (~ 7.4%) and the DBR800(-) substrate (~ -8.4%). For the former, the large difference is due to Au's large extinction coefficient (~5 at 800 nm), while for the latter the DBR800(-) substrate simply has a predicted internal intensity close to zero. The excellent agreement between $\xi$ and $\xi_{ZTA}$ indicates the internal field $\mathcal{E}_{2DM}$ inside the 2D material is to a very good approximation solely determined by the surrounding media. The result is believed to be very useful for practical applications as $\xi_{ZTA}$ can be applied to all 2D materials.

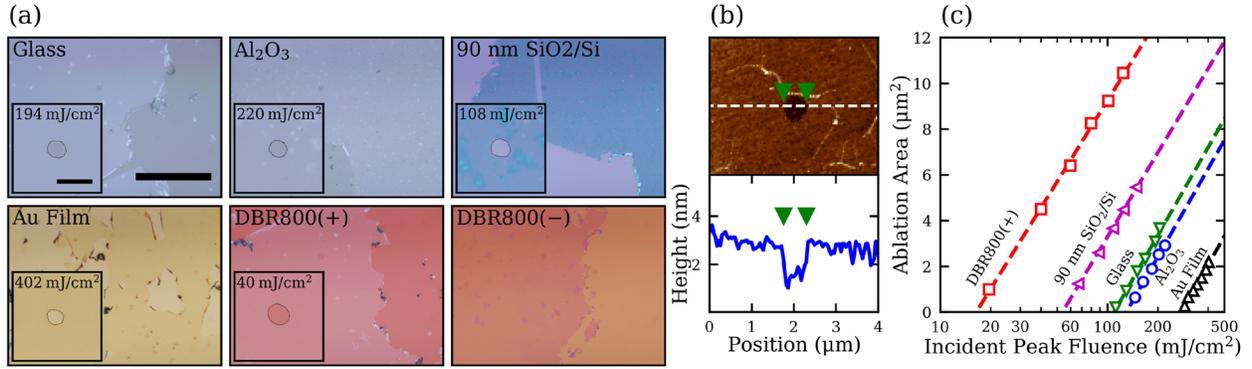

**Figure 2.** (a) Optical images of monolayer $MoS_2$ films on different substrates, demonstrating the variation in optical contrast. The scale bar is 50 μm. The inset images show ablated holes of similar ablation areas at the indicated laser fluence. The contour of these holes is outlined. The scale bar of the inset images is 4 μm. (b) AFM scan and its cross-sectional profile of a typical ablated hole of $MoS_2$ on $Al_2O_3$. (c) The ablation areas as a function of the peak fluence of the incident pulse. The intercept of the fit with the horizontal axis represents the ablation threshold, and the slope is proportional to the laser spot size.

**Intrinsic Ablation Threshold.** To experimentally investigate this etalon effect in femtosecond laser ablation of 2D materials, monolayer $MoS_2$ is used since it is one of the most widely studied TMDs, but the results here are expected to apply for all 2D materials in general. As outlined in the Materials and Methods section, monolayer $MoS_2$ films were CVD-grown on $Al_2O_3$ substrates and transferred to the all substrates used in this work (Figure 2a). A single pulse from an ultrafast amplifier operated at 160 fs and 800 nm was focused on the $MoS_2$ film using a 10x microscope objective with a 0.26 NA. The sample was translated to a fresh spot for subsequent exposures to avoid incubation effects. Figure 2a shows optical images of transferred monolayer $MoS_2$ films on various substrates where single-shot ablated holes with similar diameters are shown in the insets. The fluences ranged from 20 mJ/cm$^2$ to 400 mJ/cm$^2$, and no ablation was observed for the $MoS_2$ on the DBR800(-) substrate before the substrate itself was damaged. Overall, Figure 2a clearly demonstrates that substrates have a strong influence on the optical contrast of the films and on the ablation fluence required to make holes of similar size. Figure 2b shows an atomic force microscope (AFM) image and cross-sectional profile of a typical ablation spot in the MoS$_2$ film on Al$_2$O$_3$, indicating that material has been removed.

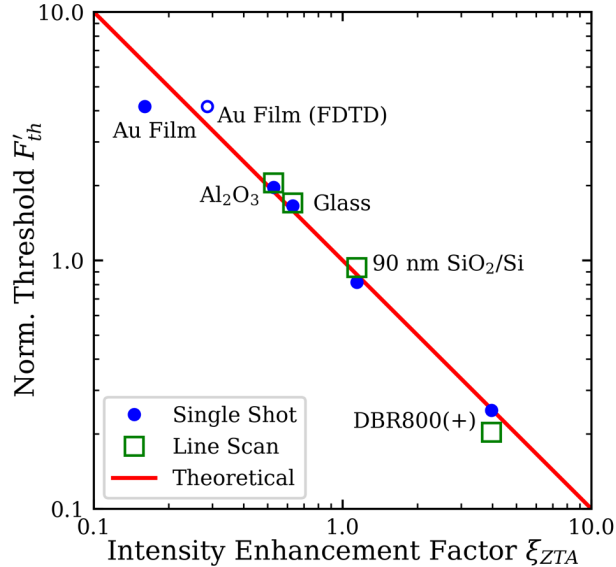

**Figure 3.** Scaling between the normalized ablation threshold and the calculated internal intensity enhancement factor at 800 nm for both single-shot and line-scan ablation. The internal intensity was calculated following the ZTA approximation for all substrate. An additional point for the internal intensity for the Au film was calculated by FDTD. The ablation threshold is normalized to the intrinsic ablation threshold $F_{th}^{int}$.

Next, to accurately measure the ablation threshold, the ablation area was measured as a function of the pulse energy, following the method outlined by Liu.[32] Figure 2c shows the ablated area as a function of peak incident fluence for different substrates. The experimentally determined threshold fluences $F_{th}$ are approximately 130, 276, 110, 54, and 16 mJ/cm$^2$, for Al$_2$O$_3$, Au film, glass, 90 nm SiO$_2$/Si, and DBR800(+), respectively. Figure 2c clearly shows that $F_{th}$ on various substrates taken by the same laser pulses can differ by an order of magnitude, confirming that substrates have a very strong influence on the laser ablation of 2D materials. According to the theory of dielectric breakdown, the onset of ablation of materials is characterized by a well-defined absorbed energy density required to break atomic bonds, corresponding to a threshold fluence. If the observed variation in $F_{th}$ is purely due to the etalon effect, $F_{th}$ should be inversely proportional to the internal intensity enhancement in MoS$_2$, that is,

$$F_{th}\xi \approx F_{th}\xi_{ZTA} = \text{constant} = F_{th}^{int}. \tag{4}$$

Equation (4) defines the intrinsic ablation threshold $F_{th}^{int}$, which is the ablation threshold fluence for a free-standing 2D material where $\xi_{ZTA}$ equals unity. $F_{th}^{int}$ is a unique threshold parameter for a 2D material that is independent of the underlying substrate. By further defining a normalized ablation threshold $F'_{th} = F_{th}/F_{th}^{int}$, equation (4) is reduced to a more compact form

$$F'_{th}\xi_{ZTA} = 1. \tag{5}$$

The experimentally determined ablation thresholds for MoS$_2$ supported by the Al$_2$O$_3$, glass, 90 nm SiO$_2$/Si, and DBR800(+) substrates in Figure 2c are fitted to $F_{th} = F_{th}^{int}/\xi_{ZTA}$ where $F_{th}^{int}$ is used as a fitting parameter (see the SI). This fit yields $F_{th}^{int} \approx 66$ mJ/cm$^2$ for monolayer MoS$_2$. $F'_{th}$ and $\xi_{ZTA}$ for various substrates are shown as solid circles in Figure 3, together with the theoretical line of equation (5). The excellent agreement for all these substrates except the Au film (to be discussed below) demonstrates that the dominating effect of these substrates in the single shot ablation of TMDs is the etalon effect, even though their thermal conductivities vary over two orders of magnitudes.[33]

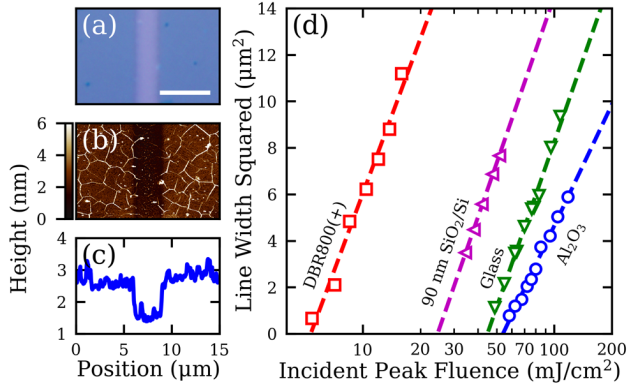

**Figure 4.** (a) An example OM image of a line patterned into a MoS$_2$ film on 90 nm SiO$_2$/Si. The scale bar is 5 μm. (b) The corresponding AFM height map to the OM image in (a). (c) An average line profile taken from the AFM height map in (b). (d) Plot of the line width squared versus the incident peak fluence for lines patterned in a MoS$_2$ on various substrates. The scan speed was set to 100 μm/s.

This result may not be too surprising, if we consider that the time scale for energy deposition by femtosecond pulses is much shorter than any phonon diffusion time such that there is little time for heat to flow into the substrates during the ablation. With high-repetition-rate femtosecond lasers, however, quasi-CW laser heating of the substrates is expected to occur such that heat transfer to the substrates may occur during the ablation. To investigate this conjecture, we conduct line-scan experiments where the MoS$_2$ film is exposed to an 80 MHz pulse train from an ultrafast oscillator while translating at a constant speed. Figures 4a-c show respectively an optical microscope (OM) image, AFM height, and AFM cross sectional profile of a line scan with a fluence of 34 mJ/cm$^2$ and a scan speed of 100 μm/s on the 90 nm SiO$_2$/Si substrate. Here, clean removal of monolayer MoS$_2$ is also observed.

Similar to the single-shot trials in Figure 2c, a line-scan ablation threshold $F_{th}$ for the MoS$_2$ film can be extracted by extrapolating the dependence of the line width squared on the peak incident fluence. Figure 4d shows the data and the fits for various substrates, taken with a fixed scan rate of 100 μm/s and a 0.26 NA focusing objective. The extracted line-scan $F_{th}$ of MoS$_2$ are 54, 49, 25, and 5 mJ/cm$^2$ for Al$_2$O$_3$, glass, 90 nm SiO$_2$/Si, and DBR800(+) substrates, respectively. Analogous to the single-shot thresholds, the line-scan thresholds are fitted to equation (4) (see the SI). This fit yields $F_{th}^{int} \approx 26$ mJ/cm² for monolayer MoS₂ at a scanning speed of 100 μm/s. The normalized thresholds $F'_{th}$ for the line-scan trials are then added to Figure 3, exhibiting again excellent agreement with equation (5). Given the thermal nature of the quasi-CW excitation, the variation of line-scan $F_{th}$ is still largely governed by the etalon effect of the substrates. We conclude that these substrates behave as very poor heat sinks for ultrafast laser ablation of 2D materials, irrespective of the substrates' thermal properties. In other words, the ablation process is adiabatic. We attribute this adiabaticity to the very low thermal boundary conductance (TBC) between MoS₂ and the substrates. Literature has reported TBC values ranging between 0.1-34 MW/m²/K for MoS₂ on SiO₂/Si substrates[34,35] and between 19-38 MW/m²/K on a sapphire substrate.[36] Additionally, mechanically exfoliated and as-grown MoS₂ monolayers on a SiO₂/Si substrate are shown to have similar TBC values. [35] Therefore, we expect that ultrafast ablation of as-grown films share the same adiabaticity as the transferred films.

Our finding that femtosecond ablation is adiabatic is in sharp contrast to the general belief that the substrates serve as a heat sink for laser processing. For example, Yoo et al. reported $F_{th} = 98$ mJ/cm² for graphene on 285 nm SiO₂/Si and $F_{th} < 43$ mJ/cm² for suspended graphene in single-shot femtosecond laser ablation.[21] They attributed this difference to the adiabatic condition of suspended graphene where heat dissipation through the substrate is forbidden. Based on our finding here, we offer an alternative interpretation. Considering the etalon effect, $\xi_{ZTA}$ are 0.2 and 1 for 285 nm SiO₂/Si (see Figure S1d) and air substrates, respectively. Based on $F_{th} = 98$ mJ/cm² for graphene on 285 nm SiO₂/Si substrate, we can estimate $F_{th} \sim 20$ mJ/cm² for a suspended graphene, which is consistent with $F_{th} < 43$ mJ/cm² reported by the authors. Moreover, the knowledge of $F_{th}^{int}$ and $\xi_{ZTA}$ (*i.e.*, Figure 1a) makes $F_{th}$ predictable for any substrate, according to equation (5). For example, given that $F_{th} = 54$ mJ/cm$^2$ (Figure 2c) and $\xi_{ZTA} = 1.14$ (Figure 1a) for the 90 nm SiO$_2$/Si substrate, the predicted threshold for the DBR800(+) substrate with $\xi_{ZTA} = 3.97$ (See Figure S1e) is $F_{th} = 15$ mJ/cm$^2$, which matches very well with the experimental threshold of 16 mJ/cm$^2$.

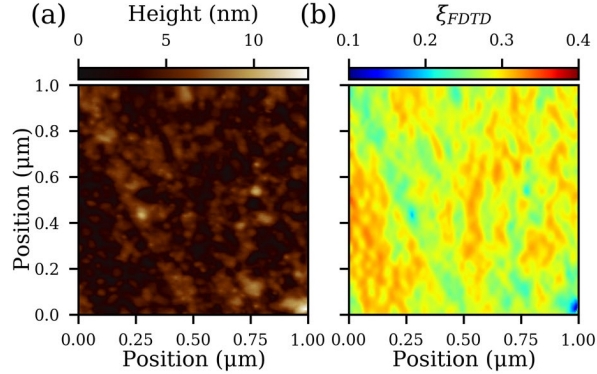

**Figure 5.** (a) AFM height scan of a 1 µm x 1 µm square of the Au surface. (b) Calculated intensity enhancement $\xi_{FDTD}$ across the simulation surface based on the AFM image in (a). See the text for details.

Among the single-shot trials (solid circles) in Figure 3, the predicted ablation threshold based on $\xi_{ZTA}$ for a smooth Au film is 40% higher than the experimental value, indicating the presence of an additional enhancement process beyond the etalon effect that increases the internal field. An AFM measurement (Figure 5a) revealed that the Au film substrate has a peak-to-peak surface roughness of 13 nm and an RMS value of 1.54 nm. This rough Au surface could lead to a local plasmonic enhancement of the incident field. Figure 5b shows a FDTD simulation of the electric field distribution at a fixed height of 0.325 nm (corresponding to half of the monolayer thickness) above the maximum height in Figure 5a. The result is only approximate, as the MoS$_2$ film may conform to the Au surface which is unaccounted in the current simulation. Additionally, the MoS$_2$ film itself is not included in the simulation to ease the computational demand and to comply with the ZTA. Nevertheless, the laterally averaged intensity enhancement factor in Figure 5b yields a much better match with $F'_{th}$ for the Au substrate, as indicated by the empty circle in Figure 3. More importantly, this result demonstrates that plasmonically active substrates could also be used to enhance the ablation of 2D materials compared to a flat metal surface. With a stronger plasmonically active substrate, even larger enhancements would be possible to further increase the ablation efficiency.

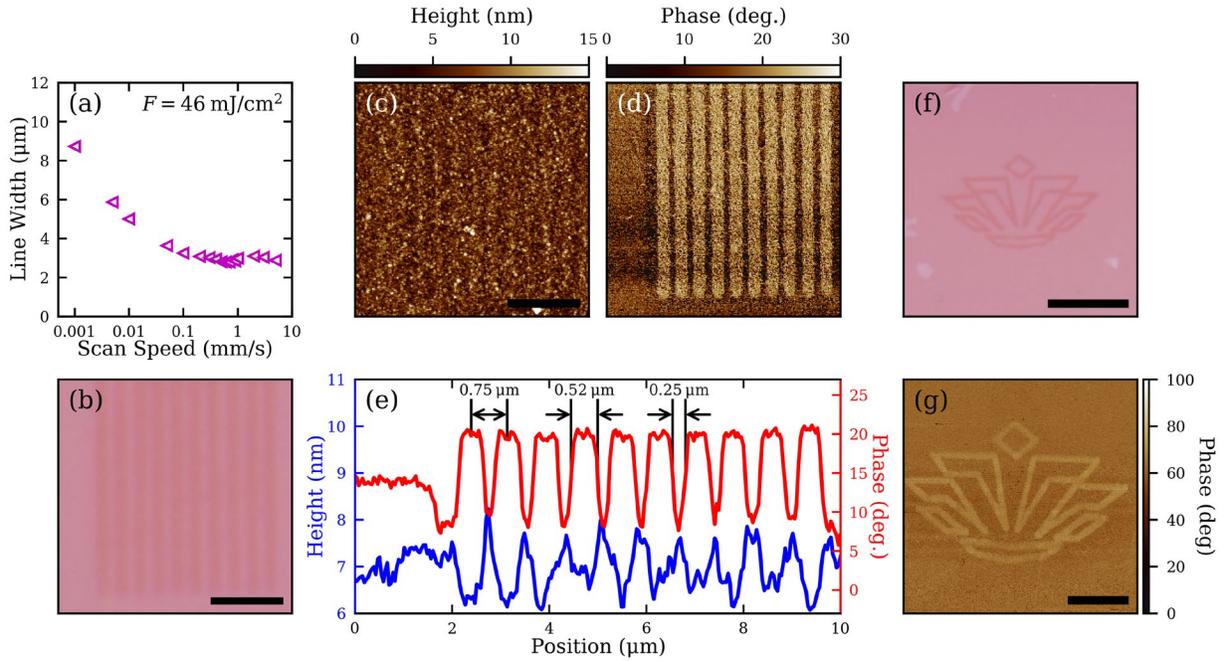

**Figure 6.** (a) A plot of the patterned linewidth in a $MoS_2$ on a 90 nm $SiO_2$/Si substrate as a function of the scan speed. (b) OM image of parallel channels patterned in a $MoS_2$ on the DBR800(+) substrate. The scale bar is 3 μm. The incident fluence was 10 mJ/cm² and the scan rate was 5 μm/s. (c) AFM height and (d) phase maps corresponding to the OM image in (b). The scale bar is 3 μm. (e) Averaged cross-sectional profiles of the AFM height and phase maps in (c) and (d). (f) OM image of the UNC Charlotte crown logo patterned into a monolayer $MoS_2$ film on the DBR800(+) substrate. The scale bar is 10 μm. The incident fluence was 10 mJ/cm² and the scan rate was 3 μm/s. (g) AFM phase map of the patterned UNC Charlotte crown in (f). The scale bar is 5 μm.

**Ultrafast Laser Patterning.** For laser patterning applications, the patterning speed and resolution are important performance metrics. Given that $SiO_2$/Si substrates are commonly used for field-effect transistors, Figure 6a shows the ablated line width in $MoS_2$ on the 90 nm $SiO_2$/Si substrate as a function of the scan rate with a constant fluence of about 46 mJ/cm² and a 0.26 NA focusing objective.[10] Selected OM images of ablated lines are shown in the SI. As the scan rate increases from 1 µm/s, the line width decreases from 8.7 µm before leveling off at 2.9 µm at 5 mm/s. The leveling off at high scan rates is due to the mechanical instability of the translation stage used here, where the stage vibrates resulting in larger widths and uneven lines (see the SI). Nevertheless, Figure 6a clearly demonstrates high-speed line patterning of TMDs. This translates into increased patterning efficiency of an ultrafast source compared to a CW source: with a scan rate of 5 mm/s, material can be removed at a rate greater than 14,000 µm$^2$/s by ultrafast lasers, whereas CW laser thinning can only pattern monolayers at a rate of 8 µm$^2$/min.[18]

To demonstrate sub-micron patterning resolution, Figures 6b-e shows an array of ablated lines in a MoS$_2$ film on the DBR800(+) substrate obtained with a 50×, 0.55 NA focusing objective. The AFM height image has poor quality due to the surface roughness of the DBR800(+) substrate (see the SI), while the AFM phase image clearly resolves the grating pattern where an average trench width of 0.52 um and ribbon width of 0.25 um are measured. To demonstrate laser micro-patterning, the UNC Charlotte crown logo was patterned into a MoS$_2$ film on the DBR800(+) substrate as shown in Figure 6f. The total size of the pattern is 20 µm and was engraved using a fluence of 10 mJ/cm$^2$ and a low feed rate of 3 µm/s to avoid skewing the pattern (see the SI). The thicknesses of the lines in the logo were found to be around 0.7 µm as measured by the AFM phase mapping. For practical applications, cost is also an important consideration. Although Figures 2c and 4d have demonstrated femtosecond ablation and patterning of MoS$_2$ on several substrates, the large field enhancement of the DBR800(+) substrate only requires pulse energies as low as 1 nJ for single-shot ablation and on the order of 100 pJ for line scans, as demonstrated in Figure 6. This pulse energy translates to an average power of 80 mW which is readily available from compact femtosecond oscillators (see the SI). With a proper design, the substrate could be engineered to enhance both the patterning process and the light-coupling performance of the resulting device. Alternatively, the patterned film can be transferred to other substrates.[19,37]

**CONCLUSION**

In conclusion, femtosecond laser patterning of monolayer MoS$_2$ was performed for the first time, where we demonstrated scan rates as high as 5 mm/s and resolutions as low as 250 nm under modest focusing conditions. We observed a nearly 20× variation in the threshold fluence for the femtosecond ablation of transferred MoS$_2$ monolayer on several substrates. This variation is attributed to the etalon effect where the substrate modulates the internal light intensity within the monolayer. An intrinsic ablation threshold $F_{th}^{int}$ is thereby introduced as a substrate-independent threshold parameter for laser ablation of 2D materials, which were found to be 66 mJ/cm$^2$ and 26 mJ/cm$^2$ for single-shot and quasi-CW ablation (80 MHz pulse train at a scanning speed 100 μm/s), respectively, for MoS$_2$. With this knowledge, we showed that the incident threshold fluence on any substrate is easily predicted. Additionally, we proved that the ablation process is adiabatic due to the very poor thermal boundary conductance between the monolayer and the substrates, which contradicts the common view that substrates serve as heat sink for laser processing. Importantly, we also introduced the zero-thickness approximation for quick and accurate estimation of the etalon effect in monolayers, which is shown to be independent of the 2D materials and applicable for any optical excitation of 2D materials beyond laser ablation. Furthermore, substrate engineering is demonstrated to enhance the ablation efficiency by 7×, enabling future patterning of 2D materials with low-power oscillators. Finally, the notion of the intrinsic threshold fluence highlights the importance of invoking the internal field instead of the incident field for studying strong-field phenomena in monolayers, including nonlinear absorption, saturable absorption, dielectric breakdown, etc., which, as it stands, also have significantly conflicting reported values, largely because they all neglect the etalon effect in their analysis.[38,39] Although transferred MoS$_2$ monolayers were studied in this work, we expect our findings can be generalized to other 2D materials, both transferred and as-grown. Our work elucidates the role of substrates and firmly establishes femtosecond laser ablation as a viable route to pattern 2D materials.

**METHODS**

**Sample Preparation.** Highly-oriented, monolayer MoS$_2$ films were grown by CVD on Al$_2$O$_3$ following the procedure outlined in reference 40.[40] All films were transferred to their host substrates which included 70 nm Au film, Al$_2$O$_3$, borosilicate glass, 90 nm SiO$_2$/Si, and two different DBR substrates. The transfer process is also outlined in reference 40.[40]

**Single-Shot Experiments.** A Coherent RegA 9000 operating at 800 nm with a pulse duration of 160 fs at the sample surface was used for all single-shot experiments. The laser was operated at a repetition rate of 307 Hz and a mechanical shutter was used to select out single pulses. Each spot on the film was only exposed to a single pulse in order to avoid incubation effects. The pulse energy was recorded with a calibrated photodiode. Optical images of the ablation features were captured using an Olympus BX51 optical microscope. The ablation areas were measured with the software ImageJ. A minimum of five ablation features were made per pulse energy and averaged for analysis.

**Laser-Patterning Experiments.** A Spectra-Physics Tsunami operating at 800 nm with a pulse duration of 210 fs and a repetition rate of 80 MHz was used for all line scan and laser patterning experiments. Sample translation and positioning was performed using an Aerotech ANT three-axis motorized translation stage. The pulse energy was simultaneously recorded using a calibrated photodiode.

# Supplemental Information:
# Ultrafast Laser Ablation, Intrinsic Threshold, and Nanopatterning of Monolayer Molybdenum Disulfide

Joel M. Solomon[1], Sabeeh Irfan Ahmad[1], Arpit Dave[1], Li-Syuan Lu[2,3], Fatemeh HadavandMirzaee[1], Shih-Chu Lin[2], Sih-Hua Chen[2], Chih-Wei Luo[2,4,5], Wen-Hao Chang[2,3], and Tsing-Hua Her[1,*]

[1] Department of Physics and Optical Science, The University of North Carolina at Charlotte, Charlotte, North Carolina 28223, United States
[2] Department of Electrophysics, National Yang Ming Chiao Tung University, Hsinchu 30010, Taiwan
[3] Research Center for Applied Sciences, Academia Sinica, Taipei 11529, Taiwan
[4] Institute of Physics and Center for Emergent Functional Matter Science, National Yang Ming Chiao Tung University, Hsinchu 30010, Taiwan
[5] National Synchrotron Radiation Research Center (NSRRC), Hsinchu 30076, Taiwan
* ther@uncc.edu

## Calculation of Internal field in 2D Materials

When light is incident on 2D materials (2DMs) such as graphene, hexagonal boron nitride, or transition metal dichalcogenides (TMDs), the field inside the 2DM can be enhanced or attenuated due to the etalon effect between the air, 2DM, and substrate. The system can be modeled as an asymmetric etalon composed of air (refractive index $\tilde{n}_0$), 2DM ($\tilde{n}_1$), and a substrate ($\tilde{n}_s$) (See Figure S1a). By considering a normally incident field $\mathcal{E}_{inc}$ and their multiple reflections from two interfaces, the electric field $\mathcal{E}_{2DM}$ at a point $x$ within the 2DM follows the standard Airy formula to be

$$\mathcal{E}_{2DM}(x) = \mathcal{E}_{inc}\tilde{t}_{01}\left(\frac{e^{i\beta_1 x} + \tilde{r}_{1s}e^{i\beta_1(2d_1-x)}}{1 - \tilde{r}_{1s}\tilde{r}_{10}e^{i2\beta_1 d_1}}\right), \qquad (S1)$$

where $\beta_j = 2\pi\tilde{n}_j/\lambda_0$, $d_1$ is the thickness of the 2DM, $\tilde{t}_{01} = 2/(1 + \tilde{n}_1)$ and $\tilde{r}_{10} = (\tilde{n}_1 - 1)/(\tilde{n}_1 + 1)$ are Fresnel coefficients, and $\tilde{r}_{1s}$ is the effective reflection coefficient of the substrate, as seen from the 2DM. The internal intensity quoted in this work is the average $|\mathcal{E}_{2DM}|^2$, calculated from equation (S1) by

$$|\mathcal{E}_{2DM}|^2 = \frac{1}{d_1}\int_0^{d_1}\mathcal{E}_{2DM}^*(x)\mathcal{E}_{2DM}(x)\,dx. \qquad (S2)$$

For single-material substrates such as Al$_2$O$_3$, cover glass, or a thick Au film, $\tilde{r}_{1s}$ is simply the Fresnel reflection coefficient:

$$\tilde{r}_{1s} = \frac{\tilde{n}_1 - \tilde{n}_s}{\tilde{n}_1 + \tilde{n}_s}. \tag{S3}$$

For SiO$_2$/Si substrates, $\tilde{r}_{1s}$ can be calculated analytically using an asymmetric etalon composed of a TMD, SiO$_2$ and Si, which yields

$$\tilde{r}_{1s} = \frac{(\tilde{n}_2 + \tilde{n}_s)(\tilde{n}_1 - \tilde{n}_2) + (\tilde{n}_1 + \tilde{n}_2)(\tilde{n}_2 - \tilde{n}_s)e^{i2\beta_2 d_2}}{(\tilde{n}_2 + \tilde{n}_s)(\tilde{n}_1 + \tilde{n}_2) + (\tilde{n}_1 - \tilde{n}_2)(\tilde{n}_2 - \tilde{n}_s)e^{i2\beta_2 d_2}}, \tag{S4}$$

where $d_2$ is the thickness of the SiO$_2$ layer and $\tilde{n}_1$, $\tilde{n}_2$, and $\tilde{n}_s$ are the refractive indices of the 2DM, SiO$_2$, and Si, respectively. $\tilde{r}_{1s}$ as a function of the thickness of the SiO$_2$ layer is shown in Figure S1c, from which the intensity enhancement factor $\xi$ based on equation (S1) is shown as the blue solid line in Figure S1d.

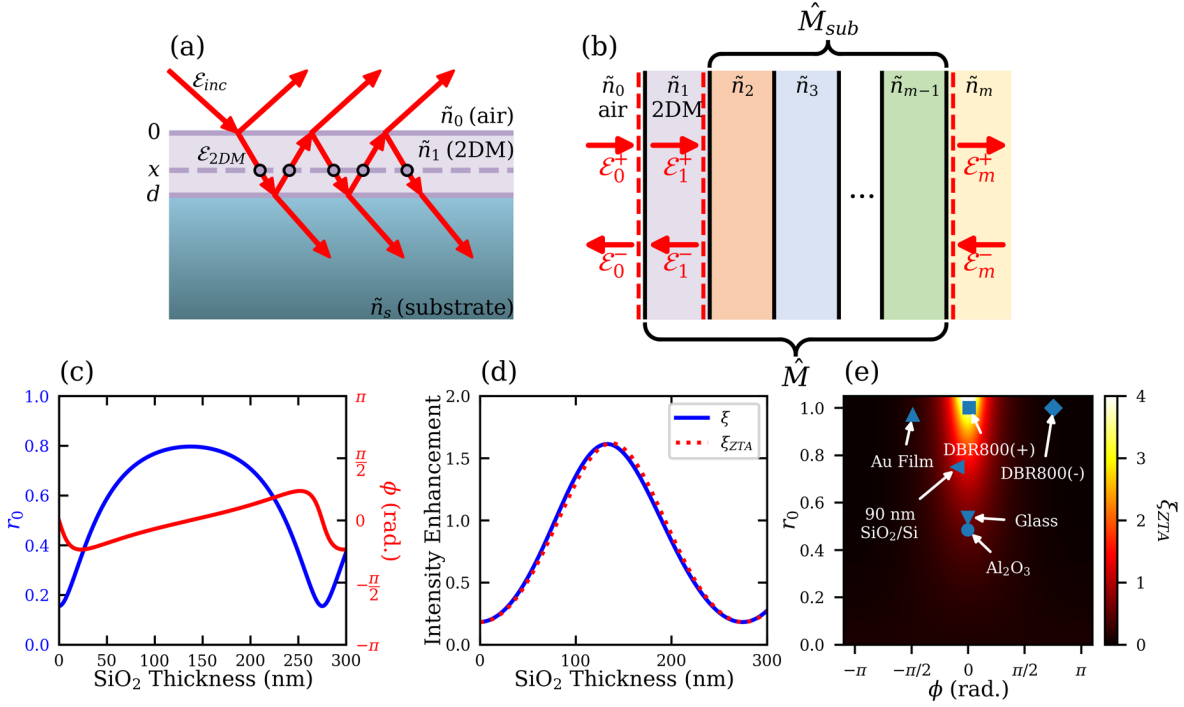

**Figure S1.** (a) Diagram for calculating the internal field strength in a 2D material. (b) Diagram for calculating the internal field in a 2D material on a stratified media based on a transfer matrix method. (c) The calculated effective reflection coefficient $\tilde{r}_{1s}$ between a MoS$_2$ and a SiO$_2$/Si substrate as a function of SiO$_2$ thickness, when excited at 800 nm. The reflection coefficient $\tilde{r}_{1s}$ is expressed as $\tilde{r}_{1s} = r_0 e^{i\phi}$ to plot in terms of its amplitude and phase. (d) The intensity enhancement factor for an MoS$_2$ film supported by an SiO$_2$/Si substrate. (e) The calculated intensity enhancement factor as a function of the effective reflection coefficient $\tilde{r}_{1s} = r_0 e^{i\phi}$. The substrates tested in the main text are marked within the plot.

For arbitrary stratified substrates, the effective reflection coefficient $\tilde{r}_{1s}$ can be calculated using a transfer matrix method (TMM). The layer indices are shown in Figure S1b, where dashed lines indicate the position to which fields are referenced. The substrate is composed of layers 2 through $m$-1, which connects the input and output fields according to

$$\begin{bmatrix} \mathcal{E}_1^+ \\ \mathcal{E}_1^- \end{bmatrix} = \hat{I}_{12}\hat{T}_2\hat{I}_{23}\hat{T}_3 \cdots \hat{T}_{m-1}\hat{I}_{m-1,m} \begin{bmatrix} \mathcal{E}_m^+ \\ \mathcal{E}_m^- \end{bmatrix} = \hat{M}_{sub} \begin{bmatrix} \mathcal{E}_m^+ \\ \mathcal{E}_m^- \end{bmatrix}, \quad \text{(S5)}$$

where

$$\hat{I}_{jk} = \text{Interface Matrix} = \frac{1}{\tilde{t}_{jk}}\begin{bmatrix} 1 & \tilde{r}_{jk} \\ \tilde{r}_{jk} & 1 \end{bmatrix},$$

$$\hat{T}_j = \text{Propagation Matrix} = \begin{bmatrix} e^{-i\beta_j d_j} & 0 \\ 0 & e^{i\beta_j d_j} \end{bmatrix}, \quad (S6)$$

$$\widehat{M}_{sub} = \text{Substrate Matrix} = \begin{bmatrix} M_{sub,00} & M_{sub,01} \\ M_{sub,10} & M_{sub,11} \end{bmatrix}.$$

In the above equation, $\tilde{r}_{jk}$ is the Fresnel reflection coefficient, $\tilde{t}_{jk}$ is the Fresnel transmission coefficient from the $j^{\text{th}}$ to $k^{\text{th}}$ medium, and $d_j$ is the thickness of the $j^{\text{th}}$ material. If the field in the final layer traveling left ($\mathcal{E}_m^-$) is zero, dividing equation (S5) by $\mathcal{E}_1^+$ yields the effective reflection and transmission coefficients of the composite substrate given by

$$\tilde{r}_{1s} = \frac{\mathcal{E}_1^-}{\mathcal{E}_1^+} = \frac{M_{sub,10}}{M_{sub,00}},$$

$$\tilde{t}_{1s} = \frac{\mathcal{E}_m^+}{\mathcal{E}_1^+} = \frac{1}{M_{sub,00}}. \quad (S7)$$

To prove that $\mathcal{E}_{2DM}^{ZTA}$ is independent of the 2DM for arbitrary stratified substrates, we note that the only dependence on the 2D material's refractive index is at the interface between layers 1 and 2. We then define a $\hat{S}$ matrix according to $\widehat{M}_{sub} = \hat{I}_{12}\hat{S}$, and the effective reflection coefficient $\tilde{r}_{1s}$ can easily be shown to be

$$\tilde{r}_{1s} = \frac{S_{10} + \tilde{r}_{12}S_{00}}{S_{00} + \tilde{r}_{12}S_{10}}, \quad (S8)$$

where equation (S8) is rigorous. To prove that $\mathcal{E}_{2DM}^{ZTA}$ is independent of the 2DM, substituting equation (S8) into equation (1) of the main text gives

$$\mathcal{E}_{2DM}^{ZTA} = \mathcal{E}_{inc}\frac{2(S_{00} + S_{10})}{(1+\tilde{n}_2)S_{00} + (1-\tilde{n}_2)S_{10}}. \quad (S9)$$

Equation (S9) offers a general proof that the $\mathcal{E}_{2DM}^{ZTA}$ is independent of the 2D material's refractive index ($\tilde{n}_1$) for any arbitrary substrate when light is at normal incidence.

For the DBR800(+) substrate with $N$ periodic bi-layers (see Figure S2a; even layers are SiO$_2$ and odd layers are TiO$_2$), the $\hat{S}$ matrix at normal incidence can be calculated analytically according to[1]

$$\hat{S} = \begin{bmatrix} S_{00} & S_{01} \\ S_{10} & S_{11} \end{bmatrix} = \begin{bmatrix} AU_{N-1} - U_{N-2} & BU_{N-1} \\ CU_{N-1} & DU_{N-1} - U_{N-2} \end{bmatrix} \tag{S10}$$

where

$$A = \frac{1}{\tilde{t}_{23}\tilde{t}_{32}} e^{-i\beta_2 d_2} e^{-i\beta_3 d_3} \left[1 - \tilde{r}_{23}^2 e^{i2\beta_3 d_3}\right]$$

$$B = \frac{\tilde{r}_{23}}{\tilde{t}_{23}\tilde{t}_{32}} e^{-i\beta_2 d_2} \left(e^{i\beta_3 d_3} - e^{-i\beta_3 d_3}\right)$$

$$C = \frac{\tilde{r}_{23}}{\tilde{t}_{23}\tilde{t}_{32}} e^{i\beta_2 d_2} \left(e^{-i\beta_3 d_3} - e^{i\beta_3 d_3}\right)$$

$$D = \frac{1}{\tilde{t}_{23}\tilde{t}_{32}} e^{i\beta_2 d_2} e^{-i\beta_3 d_3} \left(e^{i2\beta_3 d_3} - \tilde{r}_{23}^2\right)$$

$$U_N = \frac{\sin\left\{(N+1)\arccos\left[\frac{1}{2}(A+D)\right]\right\}}{\sqrt{1 - \frac{1}{4}(A+D)^2}}.$$

$$\tag{S11}$$

The $\mathcal{E}_{2DM}^{ZTA}$ for the DBR800(+) substrate at normal incident is therefore

$$\mathcal{E}_{2DM}^{ZTA} = \mathcal{E}_{inc} \frac{2(AU_{11} - U_{10} + CU_{11})}{(1+\tilde{n}_2)(AU_{11} - U_{10}) + (1-\tilde{n}_2)CU_{11}} \tag{S12}$$

where $N = 12$.

## DBR Design and Characterization

As noted in the main text, two custom designed distributed Bragg reflector (DBR) substrates are employed in this study: one of which (DBR800(+)) targets maximal intensity enhancement of 4 ($\xi_{ZTA} = 4$, $r_0 = 1$, $\phi = 0$) and the other (DBR800(-)) targets maximal intensity suppression ($\xi_{ZTA} = 0$, $r_0 = 1$, $\phi = \pi$) for 800 nm light. Both DBRs feature multiple quarter-wave stacks of a low index material $SiO_2$ ($n = 1.45$) and a high index material $TiO_2$ ($n = 2.08$) for a center wavelength of 800 nm. The design of the DBR800(+) is shown in Figure S2a, which contains 12 pairs of stacks including the $SiO_2$ as the terminating layer to produce a zero-phase shift upon reflection for total constructive interference within the 2D material. The design of the DBR800(-) is shown in Figure S2d, which contains 11 pairs of stacks plus additional $TiO_2$ as the terminating layer to produce a $\pi$-phase shift upon reflection for total destructive interference within the 2D material.

To account for manufacturing imperfections in thickness control of the DBR substrates, their reflectivities were measured and fitted to their theoretical design to extract actual layer thicknesses (Figures S2b and S2e). These extracted values are then used to calculate the theoretical reflectivity and phases (Figures S2c and S2f) and the effective reflection coefficients of the DBR substrates using equation (S8), which yield $r_0 = 1.0$, $\phi = 0.011\pi$, and $r_0 = 1.0$, $\phi = 0.76\pi$ for the DBR800(+) and DBR800(-), respectively. Measuring the actual layer thicknesses is important to best model the DBR's performance and to predict the actual field enhancement or attenuation experienced by the 2D material.

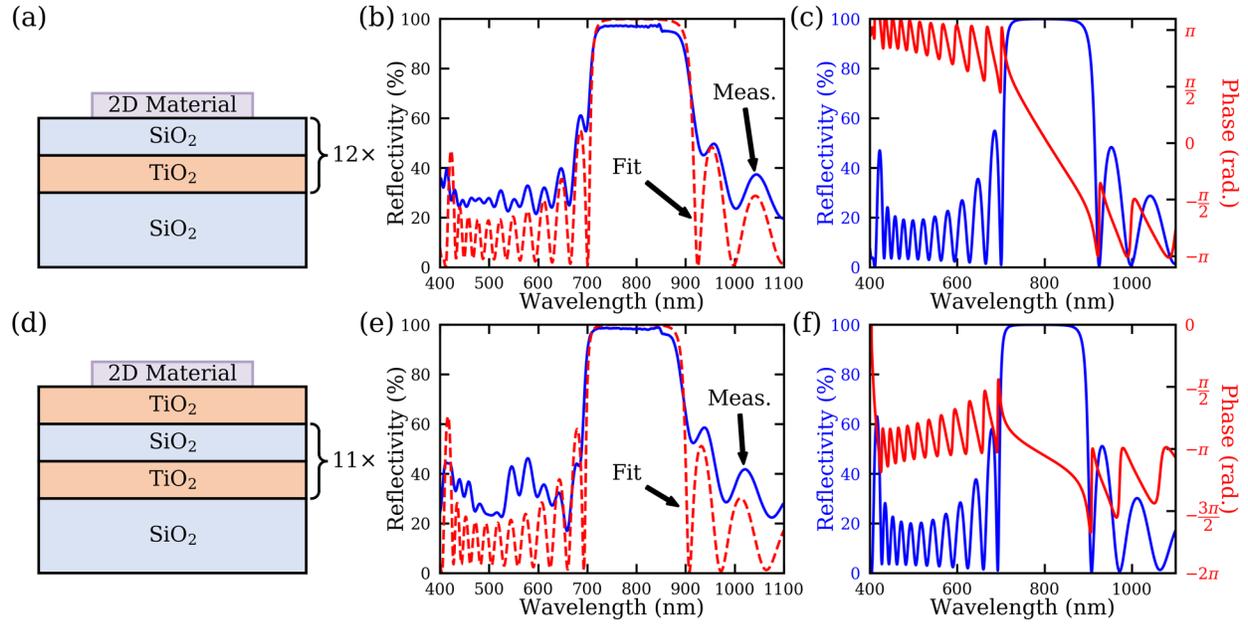

**Figure S2.** (a) Design for the DBR800(+) substrate. (b) Measured and fitted reflectivity for the DBR800(+). (c) Calculated reflectivity and phase for the DBR800(+). (d) Design for the DBR800(-) substrate. (e) Measured and fitted reflectivity for the DBR800(-). (f) Calculated reflectivity and phase for the DBR800(-).

As part of the DBR and $MoS_2$ film characterization, AFM scans of the surface were taken. Figure S3a shows an AFM height map of the edge of the $MoS_2$ film on the DBR800(+) substrate. The height of the monolayer is difficult to resolve due to the surface roughness of the DBR substrate itself. The $MoS_2$ film can be clearly resolved using the phase mapping capability of the AFM instrument as shown in Figure S3b. A significant phase change is measured when the probe is tapping the DBR surface versus the $MoS_2$ film. This capability allows one to identify when the material properties on the surface has changed. Although the DBR surface was not smooth, the surface roughness had negligible impact on the surface field enhancement as seen in the good agreement between the experimental results and the TMM calculations (Figure 3 in the main text).

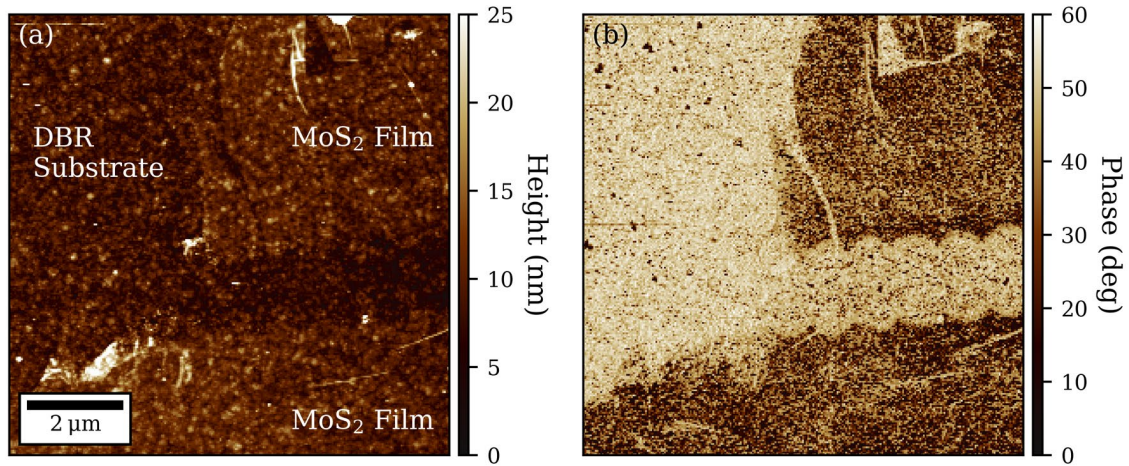

**Figure S3.** (a) AFM height map of monolayer MoS$_2$ supported by the DBR800(+) substrate. (b) AFM phase map for the same region as in (a).

# Determination of Focused Laser Spot Size, Threshold Fluence, and Intrinsic Threshold fluence

For a Gaussian spot, the diameter $D$ of an ablation feature is given by[2]

$$D_{u,v}^2 = 2w_{u,v}^2 \ln\left(\frac{E}{E_{th}}\right) = 2w_{u,v}^2 \ln\left(\frac{F}{F_{th}}\right) \tag{S13}$$

where $w$ is the laser spot radius at an intensity $e^{-2}$, $E$ is the pulse energy, $E_{th}$ is the pulse energy at the ablation threshold, $F$ is the peak fluence of the pulse, and $F_{th}$ is the peak fluence at threshold. The subscripts $u$ and $v$ represent the major and minor axes, respectively, of the laser spot profile. The peak fluence for a pulse with a Gaussian spatial and temporal profile is given by

$$F = \frac{2E}{\pi w_u w_v} = \frac{2E}{\pi w_{eff}^2} = 2F_{ave} \tag{S14}$$

where $w_{eff} = \sqrt{w_u w_v}$ is the effective laser spot radius and $F_{ave}$ is the average laser fluence. Equation (S13) consists of two expressions for pulses with elliptical spatial profiles; however, the two equations can be combined in terms of the ablation area giving

$$A = \frac{\pi}{2} w_{eff}^2 \ln\left(\frac{E}{E_{th}}\right) = \frac{\pi}{2} w_{eff}^2 \ln\left(\frac{F}{F_{th}}\right), \tag{S15}$$

where $A$ is the ablation area. This relationship also allows the determination of both the laser spot size and ablation threshold *in situ*. The effective $e^{-2}$-intensity beam radius determined from the fits were found to be 1.9 μm for the single-shot ablation trials, which agrees with the spot measured by a beam profiler within 5%. This result suggests there is little or no lateral carrier diffusion in monolayer MoS$_2$ during the time scale of the ablation. Examples of the fits for equations (S13) and (S15) are demonstrated in Figure S4 for the ablation of monolayer MoS$_2$ on borosilicate glass.

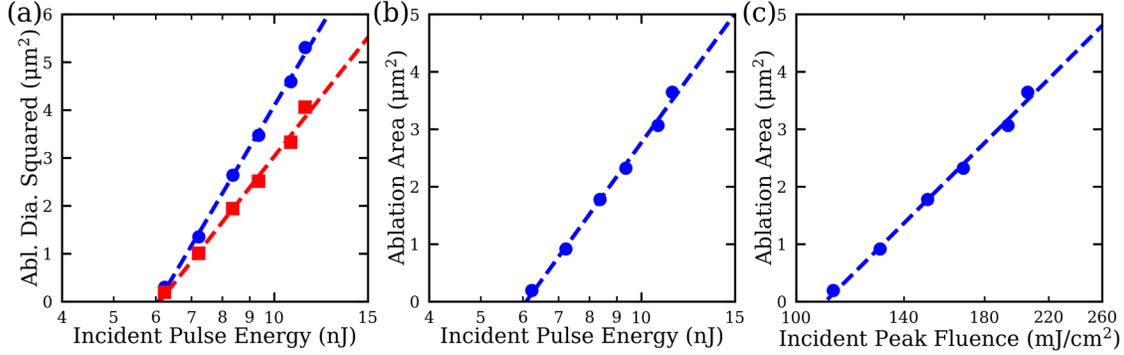

**Figure S4.** (a) Ablation diameters for $MoS_2$ on a borosilicate glass cover slip. The major and minor axes are fitted to equation (S13). (b) Ablation area for $MoS_2$ as a function of pulse energy with its fit to equation (S15). (c) Ablation area of $MoS_2$ as a function of the peak fluence of the incident pulse with its fit to equation (S15).

When determining the damage threshold with the oscillator as shown in Figure 4d of the main text, the line width is fitted to equation (S13) to extract the threshold. Although the experimental data follows the same logarithmic dependence on the fluence, care needs to be taken when extracting out the spot size. Equations (S13) and (S15) were derived under the assumption that lateral carrier diffusion is negligible. For line scans with an oscillator, material modification or breakdown is dominated by thermal mechanisms where lateral carrier diffusion may not be negligible. Additionally, line scans only scan across one axis of the laser spot which was found to be slightly elliptical. Due to these two reasons, the effective laser spot radius was determined using a CCD camera and was found to be 2.0 μm for the oscillator.

After extracting the ablation thresholds with equations (S13) and (S15) for the line scan and single shot experiments, respectively, the intrinsic ablation threshold was determined by fitting the thresholds to equation (4) in the main text. These fits are shown in Figure S5.

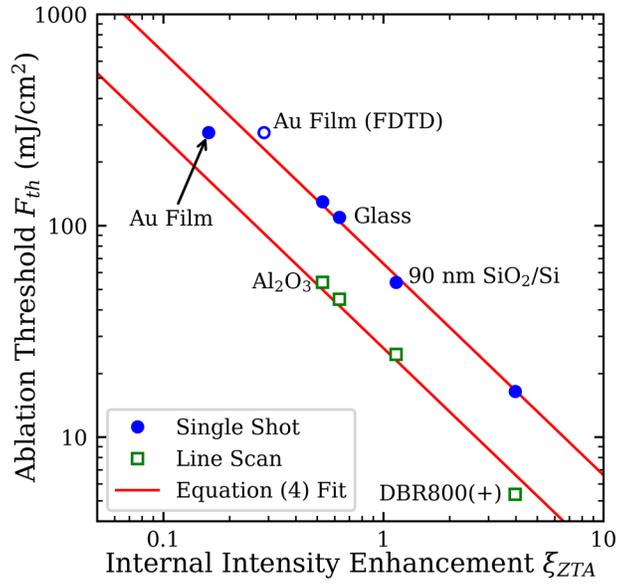

**Figure S5.** Determination of the intrinsic ablation threshold $F_{th}^{int}$ for the single shot and oscillator-based line scan experiments. For the single shot experiments, $F_{th}^{int} = 66$ mJ/cm$^2$. For the line scans with an 80 MHz oscillator, the scan rate was set to 0.1 mm/s and $F_{th}^{int} = 26$ mJ/cm$^2$.

## Ultrafast Oscillator Laser Patterning

As mentioned in the main text, laser line scans and patterning were limited by the performance of the three-axis stage used in this work. At large scan speeds, mechanical vibrations would skew lines and widen linewidths as shown in Figure S6. For the UNCC logo in Figure 6f, scan rates larger than 3 µm/s would overshoot sharp corners due the size of the logo, skewing the pattern. For the features created here, the speed was ultimately limited by the three-axis translation stage used. Galvo scanners can overcome these imitations and offer improved patterning rates and efficiencies where scan speeds reaching up 25 m/s are possible.[3]

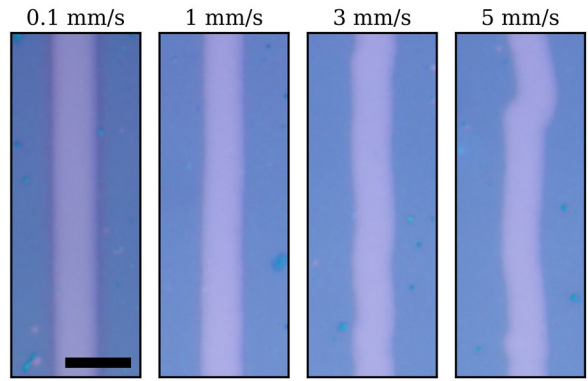

**Figure S6.** OM images of lines patterned into a $MoS_2$ film supported by a 90 nm $SiO_2$/Si substrate. The fluence was set at 46 mJ/cm$^2$. The scan rates are 0.1, 1, 3, and 5 mm/s from left to right. All images are set to the same scale with the width of the scale bar being 5 µm.

Ideally, ultrafast laser patterning of 2DMs would be conducted using oscillators due to their small form-factor and reduced cost compared to amplifiers. Fiber oscillators are especially attractive due to their superior robustness, stability, and overall ease-of-use versus their free-space counterparts. To that end, we designed the DBR800(+) substrate in order to reduce the ablation threshold of 2DMs as low as possible such that laser patterning with a commercial off-the-shelf fiber oscillator would be possible. For monolayer $MoS_2$ on the DBR800(+) substrate, single shot ablation was obtainable with 1 nJ pulse energies and laser patterning with a scan rate of 0.1 mm/s can be done with pulse energies as low as 0.37 nJ when using a 10x objective. The UNCC logo in Figure 6f in the main text was patterned with a pulse energy of 65 pJ when the pulses were focused down with a 50x objective. All-fiber, ultrafast oscillators such as erbium-doped oscillators operating at 780 nm or yitterbium-doped oscillators operating at 1030 nm typically only produce pulses that reach up to a couple of nJ. With the DBR800(+) substrate, laser patterning with these types of systems is possible. Additionally, with recent progress made in the synthesis of wafer-scale, highly-oriented $MoS_2$ films, this example further demonstrates the potential for rapid production of $MoS_2$-based devices using ultrafast laser patterning.[4] Experimentally, a high-power Ti:S oscillator (Spectra Physics Tsunami) capable of producing 10 nJ pulses at the sample was used for patterning purposes. These larger pulse energies were needed to pattern $MoS_2$ films supported by the $Al_2O_3$ and glass substrates.